# Integer and half-integer flux-quantum transitions in a niobium/iron-pnictide loop

C.-T. Chen[1*], C. C. Tsuei[1], M. B. Ketchen[1], Z.-A. Ren[2] and Z. X. Zhao[2]

[1]IBM Thomas J. Watson Research Center, Yorktown Heights, New York 10598, USA

[2]Institute of Physics and Beijing National Laboratory for Condensed Matter Physics, Chinese Academy of Sciences, Beijing 100190, P. R. China

**The recent discovery of iron-based superconductors[1-3] challenges the existing paradigm of high-temperature superconductivity. Owing to their unusual multi-orbital band structure[4-5], magnetism[6], and electron correlation[7], theories propose a unique sign-reversed *s*-wave pairing state, with the order parameter changing sign between the electron and hole Fermi pockets[8-14]. However, because of the complex Fermi surface topology and material related issues, the predicted sign reversal remains unconfirmed. Here we report a novel phase-sensitive technique for probing unconventional pairing symmetry in the polycrystalline iron-pnictides. Through the observation of both integer and *half-integer* flux-quantum transitions in composite niobium/iron-pnictide loops, we provide the first phase-sensitive evidence of the *sign change* of the order parameter in $NdFeAsO_{0.88}F_{0.12}$, lending strong support for microscopic models predicting unconventional *s*-wave pairing symmetry[9-14]. These findings have important implications on the mechanism of pnictide superconductivity, and lay the groundwork for future studies of new physics arising from the exotic order in the FeAs-based superconductors.**

---

[*] To whom correspondence should be addressed. Email: cchen3@us.ibm.com.

The new superconducting Fe-pnictides consist of two key subsets: 1) the one-layer 1111 compounds, $LnFeAsO_{1-x}F_x$, where Ln = lanthanides (including La, Nd, and Sm), with a $T_c$ ranging from 26K to 55K[1,2]; 2) the bi-layer 122 compounds, such as $(Ba,K)Fe_2As_2$, and $(Sr,K)Fe_2As_2$, with a $T_c$ up to 38K[3]. Like cuprates, the superconducting phase of Fe-pnictides emerges in close proximity to the antiferromagnetic order[6]. In addition, the low-energy electronic states are shown to derive from the conducting FeAs layers[4] (predominantly of Fe 3d character), very much like the $CuO_2$ planes in cuprates. However, striking differences do exist between the two classes of high-$T_c$ materials. For instance, the new Fe-pnictide superconductors are characterized by a multi-orbital electronic band structure and disconnected Fermi surface (FS) sheets[4,5], in contrast to the cuprates with single FS. Furthermore, the onsite Coulomb correlation is considerably weaker[7]. The similarities and disparities between the Fe-pnictides and cuprates naturally bring about the question whether the two share the same pairing mechanism. Determining the symmetry of the pair wavefunction in the new superconductors is thus of great importance for narrowing down possible microscopic theories. The existence of multiple FS sheets in these materials considerably complicates the investigations of their pairing symmetry, since various intra- and inter-FS scattering processes can participate in Cooper pairing[8-14]. Current theories of the Fe-pnictides have predicted a variety of superconducting order parameters, from spin-triplet *p*-wave pairing[15], sign-reversed extended *s*-wave ($s\pm$) pairing[9-14] to *d*-wave pairing[16,10,12]. At present, most theories favor $s\pm$ pairing when realistic material parameters are considered[9-14]. However, there is still no definitive experimental confirmation.

Here we present such an experiment designed to probe the predicted sign change of the pair wavefunction in the Fe-pnictide superconductors. Up till now, the majority of the pairing symmetry experiments have focused on the 122-type Fe-pnictide compounds[17,18]. In the case of the 1111-type superconductors, besides the complications of the FS topology, the lack of high-quality epitaxial films and large single crystals has impeded standard phase-sensitive pairing tests[19]. Almost all of the pairing experiments done on this class of materials are amplitude-sensitive only[8], except for the scanning SQUID measurements[20]. Be they thermodynamic[21-23] or tunneling-based[24], these measurements are prone to impurity contamination, surface degradation, or complications introduced by parameter fitting[8,19]. Our approach circumvents these problems by resorting to the integrity of the quantum phase in a composite Nb/NdFeAsO$_{0.88}$F$_{0.12}$ (Nb/Nd-1111) superconducting loop. We first establish supercurrent and flux quantization in the composite loop. Then, by inducing and detecting *half-flux-quantum* transitions in the loop, we establish the first phase-sensitive evidence for the sign change in the Fe-pnictide order parameter.

Figure 1 illustrates the experimental setup for observing persistent current and flux quantization. (Sample preparation and characterization are detailed in the Supplementary Information.) Our method is inspired by similar experiments on YBCO in the early days of cuprate superconductivity research[25], yet with a greater flexibility in sample exchange that enables precise calibration of the background field and against known samples, such as Nb (see Supplementary Methods). The components enclosed in the dashed red box in Fig. 1 are immersed in liquid helium with an ambient magnetic field below 50 µG. The low residual field is achieved by placing four layers of µ-metal shield around the dewar and two layers of superconducting Pb foil around the sample box. The entire system further sits in a shield room to guard against external RF contamination. Ambient white-noise level of $\leq 6\mu\Phi_0 / \sqrt{Hz}$ at ~1 $Hz$

- 3 -

measured at the dc-SQUID is routinely obtained, where flux quantum $\Phi_0 = h/2e = 2.07 \times 10^{-7} G \cdot cm^2$.

In Fig. 2, we plot the SQUID readout of the induced flux as a function of the toroidal coil current, expressed in terms of the applied magnetic flux. To conserve total flux, upon ramping up the coil current, supercurrent is induced to cancel out the applied flux threading through the Nb/Nd-1111 loop. When the induced supercurrent finally exceeds the critical current of the weakest junction in the loop, magnetic flux starts to move in or out of the loop in single or multiple units of $\Phi_0$, giving rise to the zigzag pattern in the field sweep (Fig. 2a). As we decrease the contact strength so that the junction critical current $I_c$ drops below ~20 µA, unambiguous features corresponding to single flux-quantum entries become well resolved (Fig. 2b). The persistent current in the loop implies that macroscopic quantum coherence is established across the Nb/Nd-1111 junctions. The observation of single flux-quantum entry clearly indicates that the enclosed flux is quantized and that the phase of the pair wavefunction is coherent throughout the entire loop.

To provide direct evidence of flux quantization, we present SQUID measurements of the transitions between different flux quantum states in zero fields triggered by the intermittently applied electromagnetic (EM) pulses (Fig. 3). Upon exposure to the EM disturbance, magnetic flux enters or exits the loop in quantized steps, as illustrated by the equally spaced grid lines in Fig. 3. Through comparing with the SQUID sensitivity experiment (Fig. S2) and the magnitude of flux jumps in the standard calibration Nb-Nb loop (Fig. S3), we determine the magnitude of a flux quantum jump in the Nb/Nd-1111 loop to be ~ (1.02±0.03) $\Phi_0$ (see Supplementary

Methods). As a sanity check, the same set of measurement is repeated in the open loop configuration where the junction resistance is infinite. We verify that when the circuit is open, the residual slope in the field-sweep response is consistent with the background observed in Fig. 2 and that there is *no* observable flux jump.

As demonstrated in Figs. 2 and 3, in a weak-coupling composite ring ($I_c$ below ~0.5 mA), magnetic fluxons escape or enter the loop readily at the Nb/Nd-1111 junction interface by a phase slip of $2n\pi$ (*n* is an integer), giving rise to the observed integer flux-quantum transitions. However, there is a distinct possibility that *half-integer* flux quantization can be observed in our Nb/Nd-1111 ring experiment with increasing $I_c$, if the polycrystalline Nd-1111 has a sign change (i.e. a π phase shift) in the superconducting order parameter *and* if its crystallites are properly oriented to form a π-loop. Consider a Nb/Fe-pnictide composite loop as shown in the inset of Fig. 4a. According to Refs. 26 and 27, a phase shift of π in tunneling could occur at the Nb/Nd-1111 interface. Besides, inter-grain transport between electron and hole pockets may also give rise to a π shift, although much less likely[8]. Along the current path with Nb/Nd-1111 contact areas up to ~500 μm in diameter (when $I_c$ is increased to mA range) and tip-tip separation ~ 500 μm, at least hundreds of junctions are involved, and a wide range of tunneling direction is sampled. All these significantly enhance the chance of forming π-shifted junctions. At zero magnetic field, under the experimental condition of $|I_c|L \gg \Phi_0$ where $L$ (~ 5 nH) is the self-inductance of the loop, the enclosed magnetic flux can be described by

$\Phi = n\Phi_0$,        for N = even (0-loop),

$$\Phi = (n + \tfrac{1}{2})\Phi_0, \qquad \text{for N = odd (}\pi\text{-loop)},$$

where N is the number of π phase shift around the circulating supercurrent loop[19]. The half flux-quantum becomes observable in our experiment when the large Nb loop is terminated by a Josephson junction array containing π phase shifts. In a simple model of this composite loop with a small-inductance parallel 0-π junction circuit formed at the Nb/Nd-1111 interface or within the sample, *half-integer* flux quantum is manifested in the transitions between the meta-stable quantized flux states in the big Nb loop.

We look for the signatures of such a half-flux-quantum effect by forcing the current distribution to switch between the two local-energy-minimum configurations. Toward this end, we employ a tighter coupling between the loop and the intermittently applied EM pulses. Stronger pulsed EM disturbance induces diamagnetic transient supercurrent surge that produces the driving force to propagate *half flux-quantum* across the 0-π junction array[28,29] into the Nb loop, leading to *half flux-quantum* transitions and supercurrent redistribution in the composite circuit. At the same time, we increase the contact strength to bring more Nd-1111 crystal grains in contact with the Nb tips and enhance the Josephson coupling at the Nb/Nd-1111 interface. This amplifies the probability of forming a π-shifted junction[26,27] and suppresses the integer flux quantum entries across the contact, leading to a higher occurrence of the *half flux-quantum* entries.

When $I_c$ is increased to ~5 mA, clear signals of *half-integer* flux-quantum jumps begin to show up upon EM exposure (see Figs. 4a and 4b). Four typical time traces of the transitions are presented in Figs. 4a–4d. Changes of flux states between a black grid line and a red dashed line correspond to *half-integer* flux-quantum jumps. Examples of such events are pointed out with the red arrows. As shown in Figs. 4e–4g, a total of more than one hundred *half-integer* transitions were recorded. *No* other fractional steps were observed. Taking into account of the environmental noises and drifting, we estimate that the measured *half-integer* jumps is accurate to $(0.51 \pm 0.03)\, \Phi_0$. The histograms (Figs. 4e–4g) clearly demonstrate that the probability of observing a *half flux-quantum* transition is larger for a composite loop with a higher critical current, which is consistent with the scenario of forming more π-phase-shifted junctions at the Nb/Nd-1111 contacts[26,27]. In addition, the distinctly different distribution profiles of the integer and half-integer jumps further corroborate our conjecture that they are driven by two separate physical mechanisms: the integer jumps result from the phase-slip events at the weakest contact junction, while the half-integer jumps originate from the EM-induced half flux-quantum propagation[28,29]. Using the same experimental setup, *no* half-flux-quantum jump in the Nb-Nb composite loop was detected in repeated measurements. This null result rules out incidental vortex trapping as the source of the observed $\Phi = (n + \tfrac{1}{2})\Phi_0$ jumps and reconfirms that without any π-shifted junctions, half-flux-quantum transitions can never occur. We further note that superconductor-ferromagnet-superconductor (SFS) junctions[30] are unlikely source of the observed π phase shift (see Supplementary Information for a discussion on SFS junctions and other unlikely alternative sources). Therefore, the presence of *half flux-quantum* jumps in the Nb-NdFeAsO$_{0.88}$F$_{0.12}$ loop is a direct proof of *sign change* in the order parameter of Nd-1111.

The results of our experiment are consistent with $s\pm$ and $d$-wave pairing because triplet-pairing, such as $p$-wave[15], has been ruled out by NMR experiments[23]. Among these two possibilities, $d$-wave pairing has been proven unlikely in the hole-doped 122 compounds by the $c$-axis tunneling experiments[17]. Furthermore, in $d$-wave cuprate superconductors, measurable paramagnetic Meissner effect is known to occur in the polycrystalline samples[19]. Recent scanning SQUID microscopy study on a Nd-1111 sample[20] almost identical to ours also proves against $d$-wave by the absence of such an effect. (See Ref. 8 and Supplementary Information for a discussion on the absence of paramagnetic Meissner signal in Nd-1111 and $s\pm$ pairing.) Although the aforementioned two experiments[17,20] failed to capture the $\pi$ phase shift, our observation of the *half flux-quantum* effect provides the much needed *phase-sensitive* evidence for the *sign change*, strongly supportive of $s\pm$ pairing in the 1111-type Fe-pnictide compounds.

---

1. Kamihara, Y., Watanabe, T., Hirano, M. & Hosono, H. Iron-based layered superconductor La[$O_{1-x}F_x$]FeAs (x = 0.05-0.12) with $T_c$ = 26 K. *J. Am. Chem. Soc.* **130,** 3296-3297 (2008).

2. Ren, Z.-A. *et al.* Superconductivity in iron-based F-doped layered quaternary compound $NdO_{1-x}F_x$FeAs. *Europhys. Lett.* **82,** 57002 (2008).

3. Rotter, M., Tegel, M. & Johrendt, D. Superconductivity at 38 K in the iron arsenide ($Ba_{1-x}K_x$)$Fe_2As_2$. *Phys. Rev. Lett.* **101**, 107006 (2008).

4. Singh, D. J. & Du, M. H. Density functional study of $LaFeAsO_{1-x}F_x$: a low carrier density superconductor near itinerant magnetism. *Phys. Rev. Lett.* **100**, 237003 (2008).

5. Ding, H. *et al.* Observation of Fermi-surface-dependent nodeless superconducting

gaps in $Ba_{0.6}K_{0.4}Fe_2As_2$. *Europhys. Lett.* **83**, 47001 (2008).

6. de la Cruz, C. *et al.* Magnetic order close to superconductivity in the iron-based layered $LaO_{1-x}F_xFeAs$ systems. *Nature* **453**, 899-902 (2008).

7. W. L. Yang et al. Evidence for weak electronic correlations in iron pnictides. *Phys. Rev. B* **80**, 014508 (2009).

8. Mazin, I. I. & Schmalian, J. Pairing symmetry and pairing state in ferropnictides: theoretical overview. *Physica C* **469**, 614-627 (2009).

9. Mazin, I. I., Singh, D. J., Johannes, M. D. & Du, H. M. Unconventional superconductivity with a sign reversal in the order parameter of $LaFeAsO_{1-x}F_x$. *Phys. Rev. Lett.* **101**, 057003 (2008).

10. Kuroki, K. *et al.* Unconventional pairing originating from the disconnected Fermi surfaces of superconducting $LaFeAsO_{1-x}F_x$. *Phys. Rev. Lett.* **101**, 087004 (2008).

11. Chen, W. –Q., Yang, K. –Y., Zhou, Y. & Zhang, F.-C. Strong coupling theory for superconducting iron pnictides, *Phys. Rev. Lett.* **102**, 047006 (2009).

12. Graser, S., Maier, T. A., Hirschfeld, P. J. & Scalpino, D. J. Near-degeneracy of several pairing channels in multiorbital models for the Fe pnictides. *New J. Phys.* **11**, 025016 (2009).

13. Wang, F., Zhai, H., Ran, Y., Vishwanath, A. & Lee, D. –H. Functional renormalization-group study of the pairing symmetry and pairing mechanism of the FeAs-based high-temperature superconductor. *Phys. Rev. Lett.* **102**, 047005 (2009).

14. Seo, K., Bernevig, B. A. & Hu, J. Pairing symmetry in a two-orbital exchange coupling model of oxypnictides, *Phys. Rev. Lett.* **101**, 206404 (2009).

15. Lee, P. A. & Wen, X. G. Spin-triplet p-wave pairing in a three-orbital model for iron pnictide superconductors. *Phys. Rev. B* **78**, 144517 (2008).

16. Si, Q. & Abrahams, E. Strong correlations in magnetic frustration in the high $T_c$ iron pnictides. *Phys. Rev. Lett.* **101**, 076401 (2008).

17. Zhang X., et al. Observation of the Josephson effect in Pb/Ba$_{1-x}$K$_x$Fe$_2$As$_2$ single crystal junctions. *Phys. Rev. Lett.* **102**, 147002 (2009).

18. Christianson, A. D. et al. Unconventional superconductivity in Ba$_{0.6}$K$_{0.4}$Fe$_2$As$_2$ from inelastic neutron scattering. *Nature* **456**, 930-932 (2008).

19. Tsuei, C. C. & Kirtley, J. R. Pairing symmetry in cuprate superconductors. *Rev. Mod. Phys.* **72**, 969 (2000).

20. Hicks, C. W., et al. Limits on the superconducting order parameter in NdFeAsO$_{1-x}$F$_y$ from scanning SQUID microscopy. *J. Phys. Soc. Jpn.* **78**, 013708 (2009).

21. Martin, C. et al. Non-exponential London penetration depth in RFeAsO$_{0.9}$F$_{0.1}$ (R=La,Nd) single crystals. *Phys. Rev. Lett.* **102**, 247002 (2009)

22. Hashimito, K. et al. Microwave penetration depth and quasiparticle conductivity of PrFeAsO$_{1-y}$ single crystals: evidence for a full-gap superconductor. *Phys. Rev. Lett.* **102**, 017002 (2009).

23. Nakai, Y., Ishida, K., Kamihara, Y., Hirano, M. & Hosono, H. S. Evolution from itinerant antiferromagnet to unconventional superconductor with fluorine doping in La FeAs(O$_{1-x}$F$_x$) revealed by $^{75}$As and $^{139}$La nuclear magnetic resonance. *J. Phys. Soc. Jpn.* **77**, 073701 (2008).

24. Chen, T. Y., Tesanovic, Z., Liu, R. H., Chen, X. H. & Chien, C. L. The BCS-like gap in superconductor SmFeAsO$_{0.85}$F$_{0.15}$. *Nature* **453**, 1224-1227 (2008).

25. Gough, C. E. et al. Flux quantization in a high-T$_c$ superconductor. *Nature* **326**, 835 (1987).

26. Chen, W. Q., Ma, F., Liu, Z. Y. & Zhang, F. C. $\pi$-junction to probe antiphase s-wave pairing in iron pnictide superconductors. Preprint at http://arxiv.org/abs/0906.0169v2 (2009).

27. Parker, D. & Mazin, I. I. Possible phase-sensitive tests of pairing symmetry in pnictide superconductors. *Phys. Rev. Lett.* **102**, 227007 (2009).

28. N Lazarides, N. Mobile π-links and half-integer zero-field-like steps in highly discrete alternation 0-π Josephson junction arrays. *Supercond. Sci. Technol.* **21**, 045003 (2008).

29. Chandran, M. & Kulkarni, R. V. Fractionalization of a flux quantum in a one-dimensional parallel Josephson junction array with alternating π junctions. *Phys. Rev. B* **68**, 104505 (2003).

30. V. V. Ryazanov et al. Coupling of two superconductors through a ferromagnet: evidence for a π junction. *Phys. Rev. Lett.* **86**, 2427 (2001).

**Acknowledgements** We are grateful to A. Brinkman, A. A. Golubov, D.-H. Lee, K. A. Moler, D. M. Newns, J. Z. Sun and F.-C. Zhang for stimulating discussions. We thank J. R. Rozen for technical assistance with SQUID measurements. We thank the members in Z.X. Zhao's group for assistance with sample preparation and characterization.

**Author Contributions** C.-T.C., C.C.T., and M.B.K. designed and coordinated the experiments. C.-T.C. and C.C.T. implemented the measurements, analyzed the data, and drafted the manuscript. The polycrystalline NdFeAsO$_{0.88}$F$_{0.12}$ samples were provided by Z.-A.R. and Z.X.Z.

**Competing Financial Interests statement** The authors declare that they have no competing financial interests.

**Figure 1. Schematic of the experimental setup.** The composite loop is made of a 0.75 mm-diameter Nb wire (a), polished at both ends, shaped into a 2 mm-diameter ring and pressed against the polycrystalline Nd-1111 sample (b) to form a complete circuit. Three Cu-Be spring-loaded brass screws (c) control the contact strength between the sample and the Nb tips. The distance from the center of the loop to the contact points is ~7.5 mm. The 1.6 mm-diameter Nb toroidal coil (d) applies magnetic flux to the loop from a low-noise current source via a twisted pair of leads. The Nb pickup coil (e) couples the induced flux to a commercial dc-SQUID (Quantum Design Model 50) as a magnetometer to monitor the flux states of the loop. The Nb solenoid (f) coupled to the composite loop transmits EM pulses from the external world to induce transitions between different quantized flux states.

**Figure 2. Induced flux of the composite loop with applied flux.** The horizontal axis plots the applied magnetic flux converted from the toroidal coil current. The vertical axis plots the induced flux detected by the SQUID. **a,** Typical diamagnetic response of a composite loop with a junction critical current $I_c$ ~ 50 µA. Along the sloping section, the induced supercurrent cancels out the applied flux completely to ensure flux conservation. When the applied flux is kept constant, the induced current remains at a constant value with no measurable decay, indicative of a loop resistance << $10^{-13}$ Ω. We note that when the induced supercurrent exceeds $I_c$, there remains a small residual slope in the SQUID output (see thin dashed line for a linear fit) resulting from the coupling of the SQUID pickup coil to the small leakage field of the toroidal coil. **b,** Typical response of a composite loop with $I_c$ ~ 5 µA. Single and integer-multiple flux-quantum entries upon increasing applied flux are clearly resolved. The red segments highlight single-quantum jumps, and the blue jumps describe entries of multiple flux quanta (from left to right, 4, 3, 2, 2 and 3 $\Phi_0$).

**Figure 3. Integer flux-quantum jumps in the composite Nb-NdFeAsO$_{0.88}$F$_{0.12}$ loop.** Here we plot the transitions of the magnetic flux states of the composite loop upon pulsed EM excitations in the weak-coupling limit ($I_c$ below ~0.5 mA). Three runs of inter-flux-state transition measurements have been recorded, with contacts formed at different locations on the sample surface and of varying loop critical currents in the sub-mA range. After calibrating against the standard Nb/Nb loop, the unit of a flux quantum jump in the Nb/Nd-1111 loop is determined to be ~ (1.02±0.03) $\Phi_0$, taking into account of the systematic error and the statistical variations.

**Figure 4. Half-integer flux-quantum jumps in the composite Nb-NdFeAsO$_{0.88}$F$_{0.12}$ loop.** The spacing between the adjacent black solid lines and red dashed lines is ½ $\Phi_0$. The time interval between two ticks is 50 seconds. The red arrows label the half-integer flux-quantum jumps. **a, b,** Representative data taken on a composite loop with $I_c$ ~ 5mA. Inset of Figure 4a illustrates schematically that the supercurrent distribution in the composite loop passes through a large number of grain-boundary junctions with a wide range of incident angles in the polycrystalline NdFeAsO$_{0.88}$F$_{0.12}$. The realistic current flow pattern is much more complicated than depicted. **c, d,** Representative data taken on a composite loop with $I_c$ ~ 15mA. Three sets of half-integer-transition data have been recorded with contacts formed at different locations on the sample surface and of varying $I_c$ in the mA range. **e-g,** Histograms of integer vs. half-integer transitions upon *pulsed* EM excitations. We note that the width of each bin does *not* reflect the error-bar of the measured flux jumps (which is about ±0.03 $\Phi_0$).

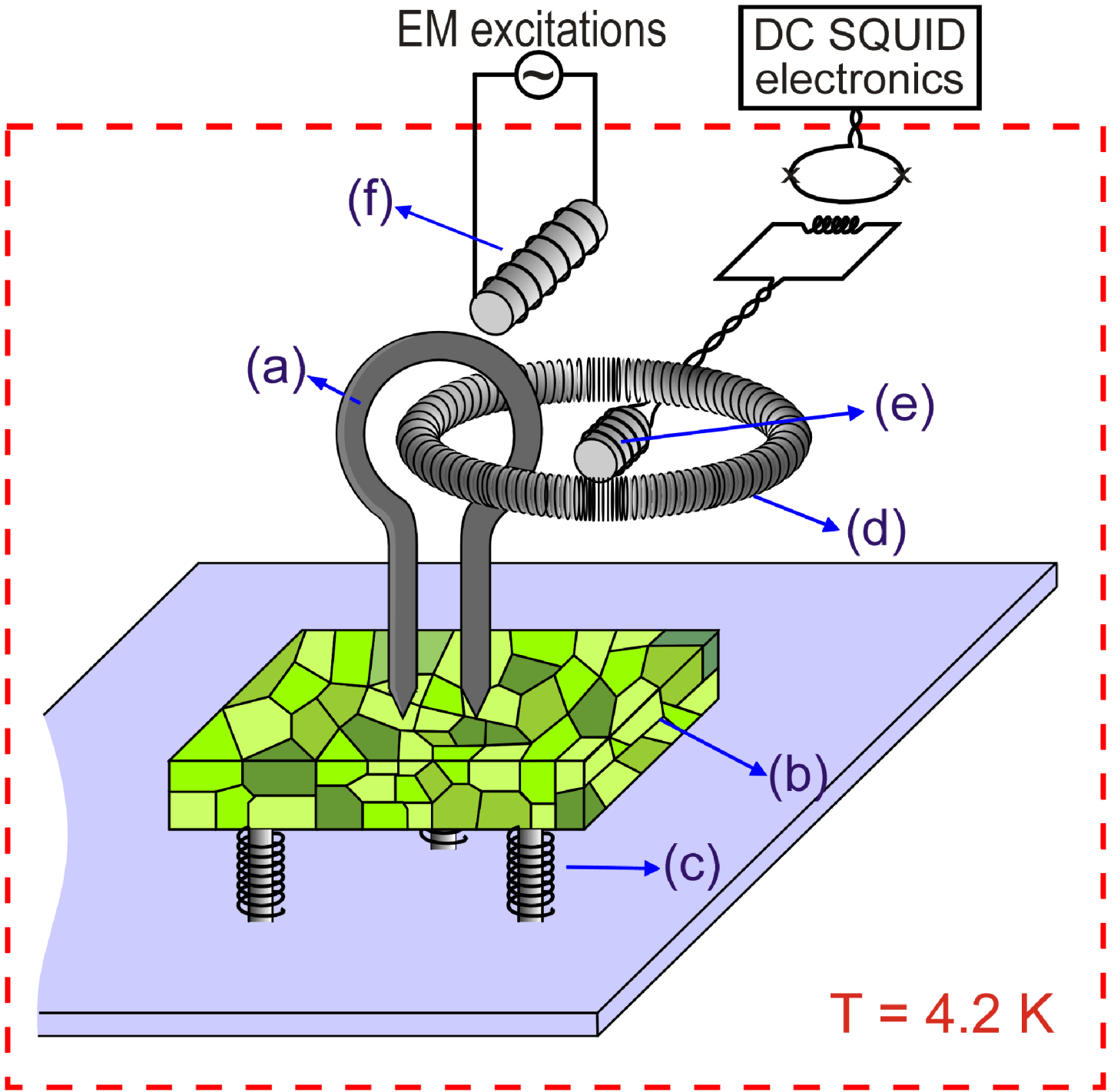

Fig. 1

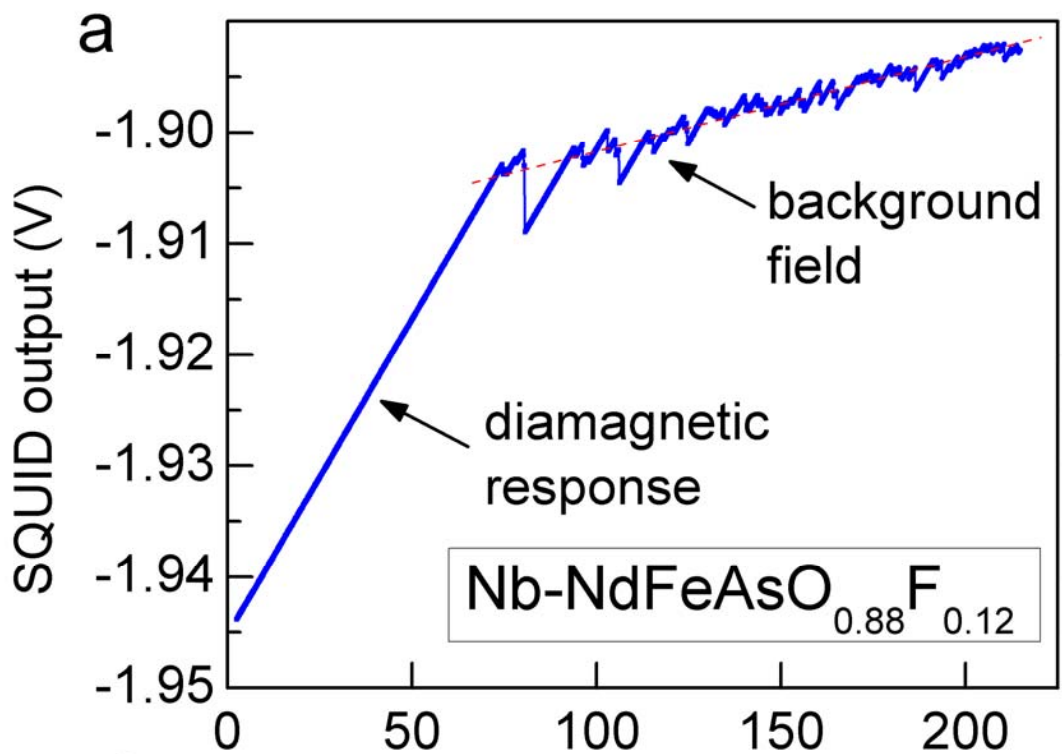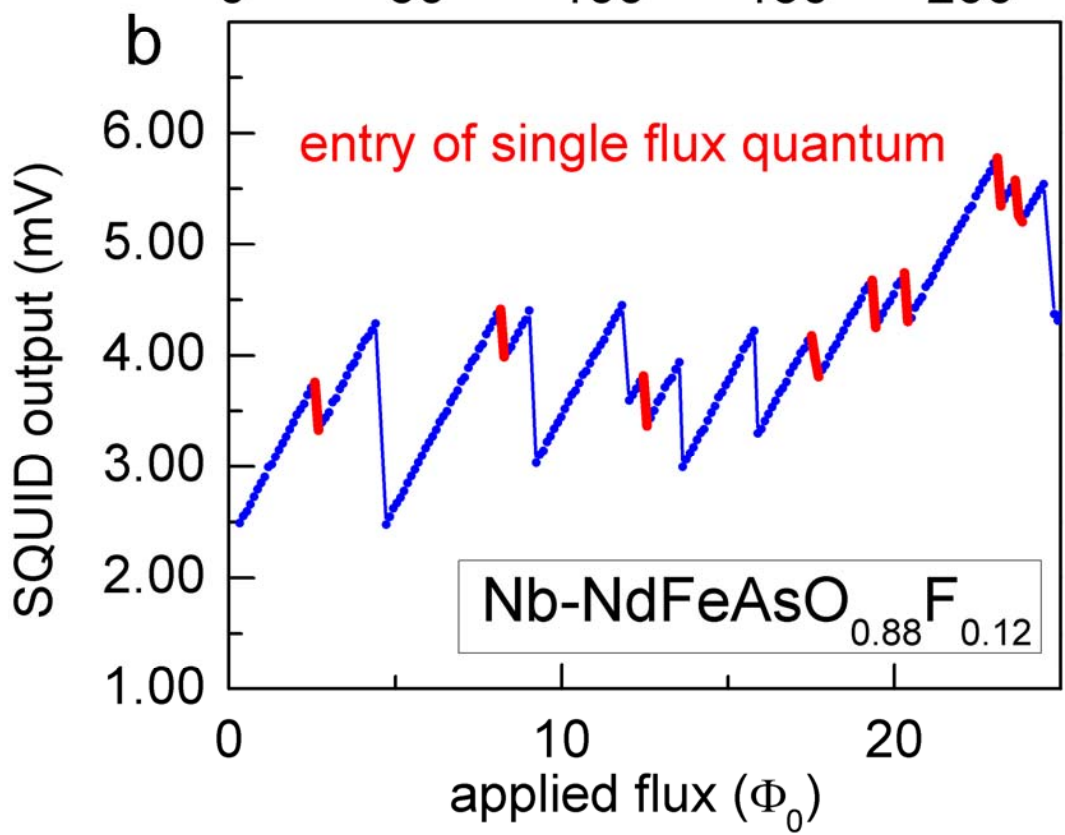

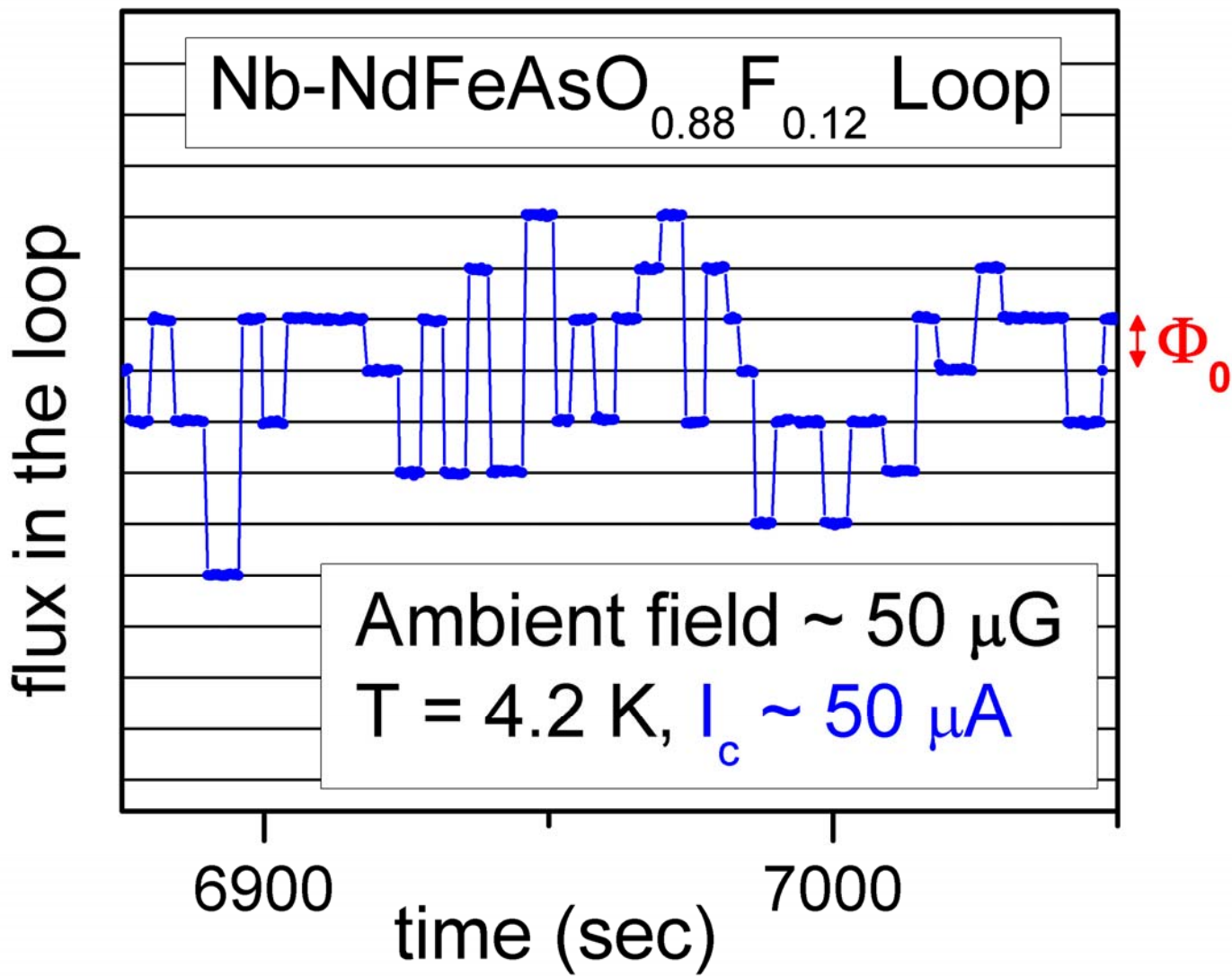

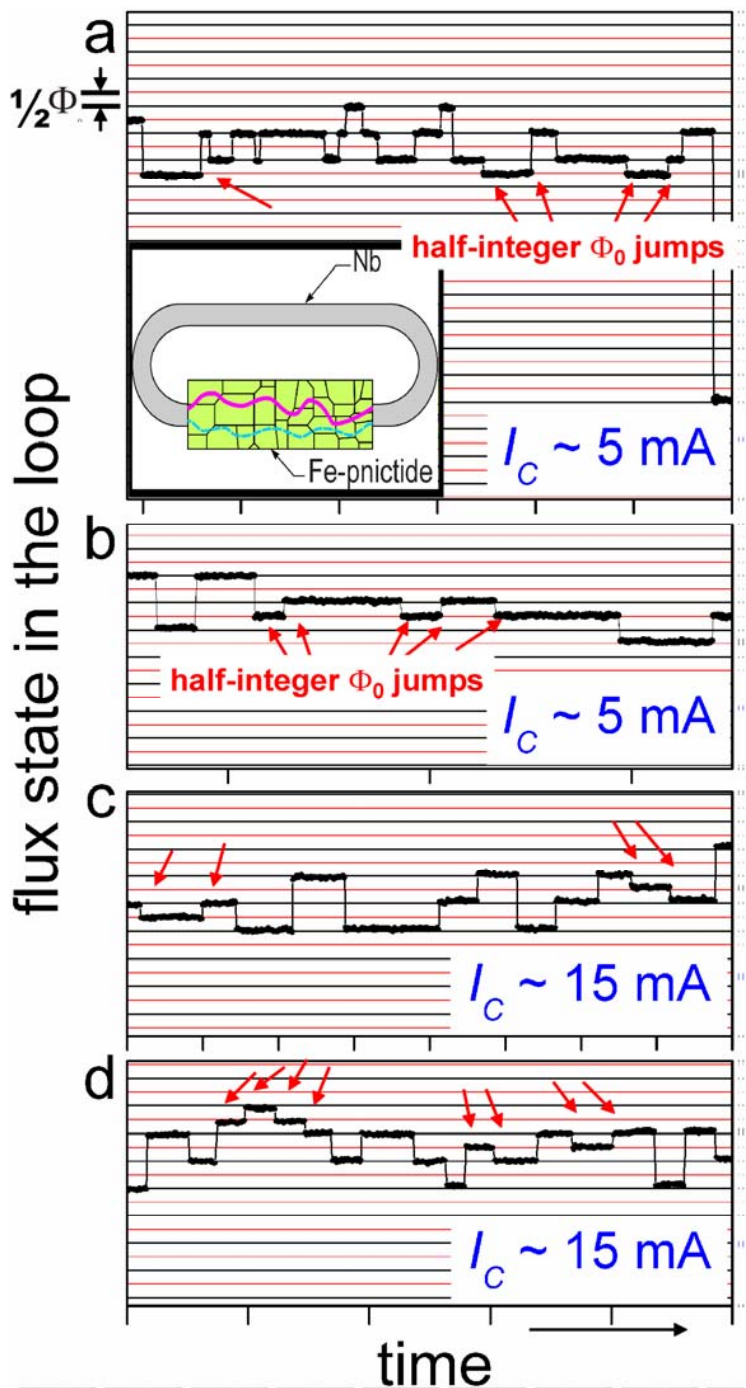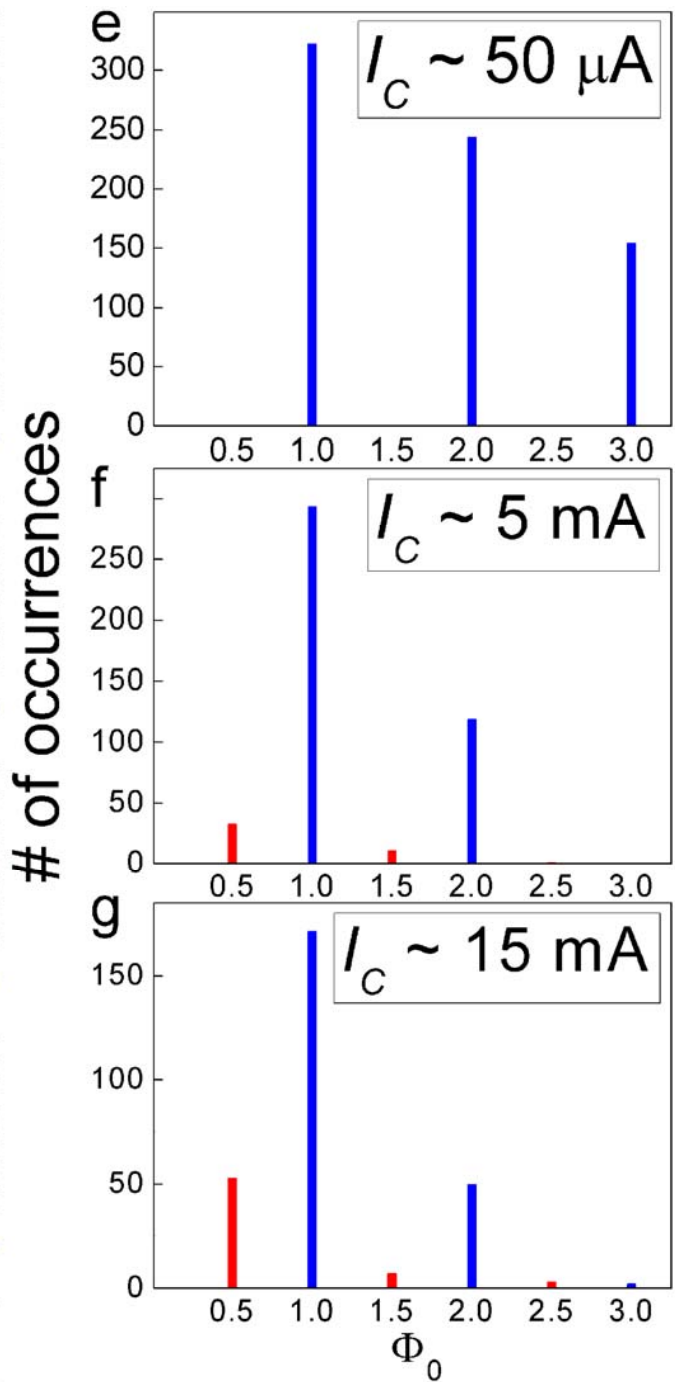

# Supplementary Information: Integer and half-integer flux-quantum transitions in a niobium/iron-pnictide loop

C.-T. Chen, C. C. Tsuei, M. B. Ketchen, Z.-A. Ren, and Z. X. Zhao

## I. Supplementary Methods:

**A. Sample preparation and characterization**

The polycrystalline $NdFeAsO_{0.88}F_{0.12}$ (Nd-1111) samples measured in the present work were synthesized by solid-state reaction at high temperatures and under high-pressure. Powders of NdAs (pre-sintered), Fe, $Fe_2O_3$, and $FeF_2$ were mixed together according to the nominal chemical stoichiometry of $NdFeAsO_{0.88}F_{0.12}$, ground thoroughly, and pressed into small pellets under inert gas atmosphere. The pellet was sealed in boron nitride crucibles and sintered in the graphite tube of the high-pressure synthesis apparatus. After the pressure was increased to 6 GPa, the temperature was increased to 1250 °C and maintained for two hours. The pressure was released after the sample fully cooled down. Details of the sample preparation can be found in the paper by Ren *et al.* [S1]. The Nd-1111 samples are characterized by a sharp normal-to-superconducting phase transition at 43K, as demonstrated by the low-field magnetic susceptibility and dc-resistivity measurements shown in Fig. S1.

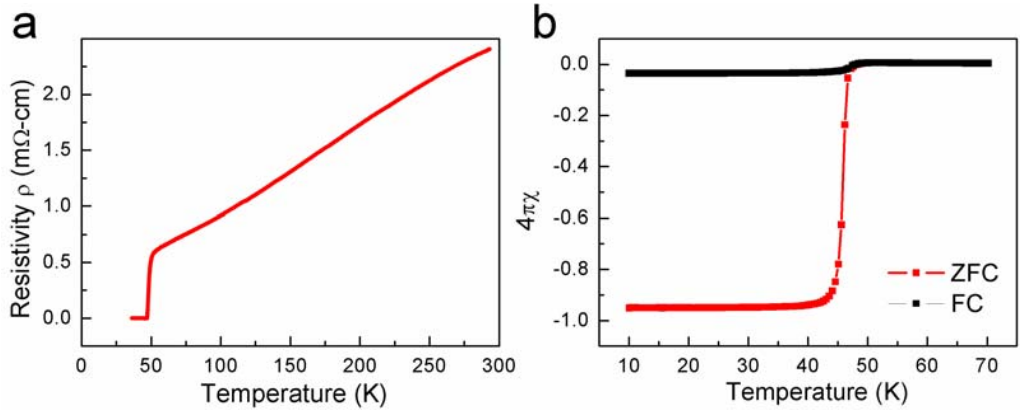

**Fig. S1. (a) Temperature dependence of the resistivity of $NdFeAsO_{0.88}F_{0.12}$.** The data were obtained with a standard four-probe method. **(b) Temperature dependence of the low-field magnetic susceptibility of $NdFeAsO_{0.88}F_{0.12}$.** The susceptibility measurements were performed on a Quantum Design MPMS XL-1 system during the warm-up period under fixed magnetic field (1 Gauss) after zero-field cooling (ZFC) and field cooling (FC) process.

The global critical current density ($J_c$) at 5K is ~ 2000 A $cm^{-2}$ [S2], about two to three orders of magnitude lower than the intra-grain $J_c$, very much like the cuprates [S3]. Furthermore, Ref. [S4] presents scanning SQUID microscopy evidence of the flux trapped in the grain boundaries. Many of the observed vortices are elongated along one axis (exceeding the resolution of 6 μm by a significant amount) and resolution-limited along the other axis, suggesting that the nature of the grain-boundary junctions in this material is similar to that in the polycrystalline cuprates.

A recent combined micro-structural and magneto-optical study on a polycrystalline sample of $NdFeAsO_{0.94}F_{0.06}$ prepared under identical synthesis conditions has revealed that the Nd-1111 sample is an assembly of randomly oriented plate-shaped grains with an average size of ~ 7x2.8 μm$^2$ and a thickness of about 5 μm [S2]. Accordingly, supercurrent passes through tens to hundreds of grains to complete the circuit in the composite loop. Among the interfaces between grains, some are clean and well connected with a distribution of $I_c$, while others contain impurity phases serving as flux pinning sites. While the supercurrent mostly flows through the clean grain boundaries, some of the inter-grain current paths are obstructed by such extrinsic defects. This non-uniform supercurrent flow distribution observed in Nd-1111 samples sets the stage for our flux quantization experiment described in this paper.

**B. Fine calibration against the Nb-Nb composite loop**

The exact value of the measured flux quanta in the Nb-$NdFeAsO_{0.88}F_{0.12}$ loop is determined in two steps. First, we follow Ref. [S5] and use the current-sweep data in Fig. 2 for a rough estimate. Along the sloping section, the total magnetic flux threading through the loop remains constant. Therefore, the applied flux completely determines the induced flux. We obtain the nominal value of the induced flux at a given applied current using the measured coil diameter (1.6±0.1) mm and turn spacing (0.099±0.005) mm. The SQUID response to an induced flux of $\Phi_0$ can then be read off from Fig. 2. We note that besides the induced diamagnetic signal, a small background field from the toroidal coil also couples to the SQUID, giving rise to the residual slope in the readout. After accounting for this background coupling, we estimate the magnitude of the single flux-quantum jumps in Fig. 3 to be $(1.15 \pm 0.20)\ \Phi_0$. The error-bar reflects the uncertainty in estimating the self-inductance of the toroidal coil. In reality, the measured signal variation of a flux quantum is much smaller ($\leqslant \pm 3\%$).

For a better calibration of the SQUID sensitivity, we replace the $NdFeAsO_{0.88}F_{0.12}$ sample with a piece of pure Nb sample with similar dimensions and repeat the same set of experiments. Comparing the SQUID data of the Nb-Nb loop (see Figs. S2 and S3) with Figs. 2 and 3, we verify that the diamagnetic slope, the background pickup, and the signal corresponding to single flux-quantum jumps are consistent to within ~ ±3%.

Figure S2 plots the field-sweep data of the two different composite loops. The SQUID output of the Nb-Nb loop is plotted in blue, and that of the Nb-$NdFeAsO_{0.88}F_{0.12}$ loop is plotted in red. The linear fits of the two datasets are consistent to within ±3%. Therefore, given the same applied current, the SQUID readings of the induced flux and the background coupling are nearly identical. Figure S3 shows the flux state transitions of the two composite loops upon intermittent electromagnetic excitations. The two sets of data are overlaid for comparison ***without scaling***. It clearly demonstrates that the flux jumps of the two loops are quantized in the same unit. After taking account of the slight differences in field-sweep slopes and averaging over multiple datasets, and knowing that a flux quantum in an all-Nb loop is exactly $\Phi_0$, we determine that the value of a single flux quantum jump in the Nb-$NdFeAsO_{0.88}F_{0.12}$ loop to be ~ $(1.02 \pm 0.03)\ \Phi_0$.

Lastly, we use the calibrated SQUID flux sensitivity to calculate the self-inductance of the toroidal coil for precise current-to-flux conversion in Fig. 2.

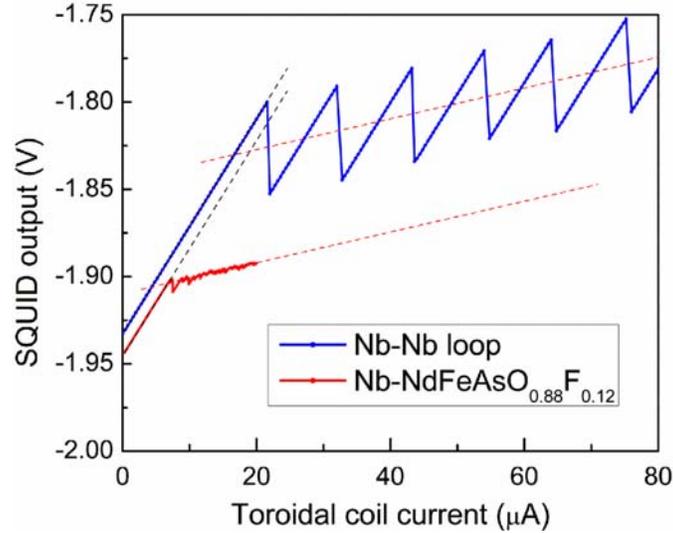

**Fig. S2. SQUID output signals of the induced flux in the composite loops.** Data of the Nb-NdFeAsO$_{0.88}$F$_{0.12}$ loop are shifted by -4 V for better contrast. The black dashed lines are linear fits to the diamagnetic sections of the data. The red dashed lines are linear fits to the background pickup from the toroidal coil. We note that the vertical jumps in the Nb-Nb loop dataset correspond to ~100 $\Phi_0$.

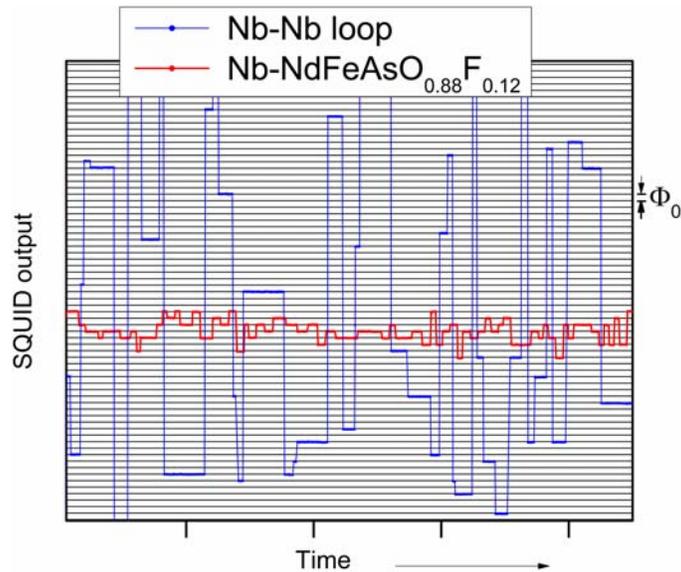

**Fig. S3. Flux state transitions of the composite Nb-Nb and Nb-NdFeAsO$_{0.88}$F$_{0.12}$ loops.** In order to better resolve single flux-quantum transitions, the electromagnetic excitations applied to the Nb-NdFeAsO$_{0.88}$F$_{0.12}$ loop were attenuated to suppress the large flux jumps seen in the Nb-Nb loop.

### C. Advantage of the composite-loop configuration

Our design has several key features. First of all, using point-contact junctions, we bypass the intricacy of making planar junctions on the 1111 Fe-pnictides which have proven to be extremely difficult to fabricate. Secondly, the spring-loaded mechanism allows us to vary the tunneling barrier in a controlled manner. This is crucial for observing single fluxon entry upon current ramping and EM excitations. More specifically, adjusting the contact strength through the spring-loaded screws enables the control of the junction critical current, which has been demonstrated as a reliable way to characterize the flux dynamics and the magnetic state transition behaviour of the composite superconducting loop. Lastly, it is very easy to exchange samples ***without*** altering the experimental condition and the background pickup. Thus, our experiment allows for very precise calibration of the SQUID output corresponding to single flux-quantum jumps via comparison against single fluxon entries in Nb-Nb composite loops.

### D. Other technical details

**a.** The residual magnetic field is measured using a fluxgate magnetometer to be less than 50 μG at room temperature. With *in situ* degaussing, our setup is capable of achieving a residual field of 3-6 μG, as reported in Ref. [S6]. At 4.2K, the superconducting shield further suppresses the magnetic fluctuations arising from the mechanical displacement and thermal drifting between the sample box and the μ-metal shield. However, the vibrations and drifting inside the sample box cannot be shielded away and thus contribute to the residual noise and signal drifting in the data.

**b.** The 1/f noise knee lies at < 1 Hz. The quoted noise value ~ $6\mu\Phi_0/\sqrt{Hz}$ is the white-noise level at the SQUID electronics output measured by an HP spectrum analyzer. The reported data in Figs. 2-4 are readings of the SQUID output taken by an Agilent voltmeter in the high-resolution mode with a signal integration time ~ 1 sec. ***There is no further averaging in the data recording.***

**c.** The conversion factor for the Quantum Design SQUID is ~0.785 V per $\Phi_0$ threaded through the SQUID. The SQUID voltage output per $\Phi_0$ threaded through the composite loop is much smaller (~0.5 mV per $\Phi_0$), and it varies sensitively with the placement of the pickup coil and the loop geometry. We have kept the experimental configuration as constant as possible when exchanging samples. The magnitude of the integrated magnetic noises, about $\pm 0.05\Phi_0$ threaded through the loop, can be read off directly from the voltmeter since the noise is dominated by the low-frequency 1/f component.

## II. Supplementary Discussion:

### A. Unlikely alternative sources of the half-Integer jumps

Here we elucidate how other explanations of the observed half-integer jumps can be ruled out.

**a. Flux trapping:** We are sure that flux trapping is ***not*** the source of the observed half-integer jumps for the following reasons. First, it would have to be a great coincidence if the flux trapping site on the niobium loop results in consecutive partial vortex jumping signals that equal to *two successive half-flux-quantum transitions*. More importantly, when the Nd-1111 sample is replaced with a pure niobium sample, *we **never** detect any half-flux-quantum transition signals in the resulting Nb/Nb loop using the same setup*, no matter how we vary the junction critical current. Only integer jumps are observed. This null result definitely rules out flux trapping on the niobium wire.

We further note that trapping on the polycrystalline Nd-1111 sample can also be ruled out, since it is located far away from the pickup coil. Flux entry via an intermediate trap site on the sample would give rise to two successive highly-asymmetric unequal fractional jumps, instead of the observed succesive half-integer jumps.

**b. Superconductor-Ferromagnet-Superconductor (SFS) junctions as a source of $\pi$-phase shift in the Nb / Nd-1111 loop:** A combined microstructural and magneto-optical study [S2] of the current flow in polycrystalline Nd-1111 (almost identical to the samples used in this work and also in the study by Moler's group [S4]) revealed that the impurity phases (amorphous FeAs phase or insulating $Nd_2O_3$) are located mainly on the *vertices* or the *void* between superconducting grains, and some along the grain boundaries. The latest chemical analysis of a closely related Sm-1111 polycrystalline sample, using energy-dispersive X-ray spectroscopy, further reveals that the poly-crystallites are surrounded by thin layers of **antiferromagnetic $Fe_2As$** which form the junction barrier (see page 5 of Ref. [S7]). In other words, the grain-boundary junctions between crystallites *cannot* be SFS junctions. Furthermore, to achieve a $\pi$-phase-shift across the junction demands a well-engineered SFS tri-layered structure that satisfies an *exact combination of requirements* on the ferromagnetic-layer thickness, exchange energy, and so on [S8]. Therefore, it is very unlikely that an incipient SFS junction can fortuitously form a $\pi$-junction and cause the observed half-flux-quantum jumps.

Regarding the sporadic contamination from the minute traces of paramagnetic/ferromagnetic impurities, such as Fe, (if exists) at the grain-boundaries, it is very unlikely that the tiny amount of isolated magnetic impurities if exists could cause a net $\pi$-flip scattering. Furthermore, we polish and thoroughly clean the surface of our sample with lint-free wipers and pure IPA/ethanol before making the Nb/Nd-1111 contacts. Trapping magnetic impurities at the contact interface for all three runs of experiments is also very unlikely.

**c. Fractionalization of magnetic flux in a multi-band superconductor:** This scenario can be ruled out for the following reasons. First, the postulated fractional flux in a two-band superconductor can in theory take on *any fraction* of a flux quantum (see for example, Ref. [S9].) It *cannot* explain why *only the half-flux quantum jumps are observed* in multiple experimental runs. Secondly, the fractionalized phases are unstable thermodynamically. In the low-temperature limit, fractional vortices are always *confined* in bulk samples [S10]. This implies that only ordinary *integer* flux quanta can be observed if there is *no* sign change in the superconducting order parameter, as demonstrated by the null results in the two-band superconductor $MgB_2$.

*In summary, the observed half-integer flux-quantum effect is not an experimental artifact, nor can it be attributed to contaminations. Therefore, the most natural explanation of the half-integer jumps is the sign change in the Nd-1111 order parameter.*

**B. Absence of paramagnetic Meissner effect in an s±-wave superconductor**

Here we explain why the absence of paramagnetic Meissner signals in scanning SQUID microscopy [S4] is consistent with our conclusion of *s*±-wave pairing symmetry.

In contrast to the case of *d*-wave cuprates, the probability of encountering a π-phase-shifted grain-boundary junction in a polycrystalline sample is extremely small for an *s*± superconductor with the band structure of Nd-1111, as pointed out in Ref. [S11]. Consequently, the probability of finding a spontaneous half flux quantum is slim, and the resulting paramagnetic Meissner signal may be below the sensitivity of scanning SQUID microscopy, consistent with the result of Ref. [S4].

On the other hand, the Nb/Nd-1111 interface in our setup provides a second venue for the formation of π phase shift [S12, S13] originating from the intrinsic sign change in *s*± pairing. This gives rise to the observed half-flux-quantum effect in our experiment that cannot be directly accessed by the scanning SQUID microscopy measurements [S4].

## III. References

[S1] Z.-A. Ren *et al.*, *Europhys. Lett.* **82,** 57002 (2008).
[S2] F. Kametani *et al.*, *Supercond. Sci. Technol.* **22**, 015010 (2009).
[S3] A. Yamamoto *et al.*, *Supercond. Sci. Technol*. **21**, 095008 (2008).
[S4] C. W. Hicks *et al.*, *J. Phys. Soc. Jpn.* **78**, 013708 (2009).
[S5] C. E. Gough *et al.*, *Nature* **326**, 835 (1987).
[S6] S. Bermon, P. Chaudhari, C. C. Chi, C. D. Tesche, C. C. Tsuei, *Phys. Rev. Lett.* **55**, 1850 (1985).
[S7] S. Sunna *et al.*, Preprint at http://arxiv.org/abs/0909.3004 (2009).
[S8] V. V. Ryazanov *et al.*, *Phys. Rev. Lett.* **86**, 2427 (2001).
[S9] J. Goryo, S. Soma and H. Matsukawa, *EPL* **80**, 17002 (2007).
[S10] L. F. Chibotaru, V. H. Dao and A. Ceulemans, *EPL* **78**, 47001 (2007).
[S11] I. I. Mazin, I. I., J. Schmalian, *Physica C* **469**, 614 (2009).
[S12] W. Q. Chen, F. Ma, Z. Y. Liu, F. C. Zhang, preprint at http://arxiv.org/abs/0906.0169v2 (2009).
[S13] D. Parker, I. I. Mazin, *Phys. Rev. Lett.* **102**, 227007 (2009).